\documentclass[conference,usenatbib]{basi}
\usepackage{lmodern}
\usepackage{lscape}
\usepackage{float}
\usepackage[T1]{fontenc}
\usepackage[british]{babel}
\usepackage[varg]{txfonts}
\usepackage{epsfig}          
\usepackage{graphicx}        
\usepackage{dcolumn}
\begin{document}
\title[Interplanetary electric potential and energy coupling]{Fluctuations in the interplanetary electric potential and energy coupling between the solar wind and the magnetosphere}
\author[Badruddin \& O.~P.~M. ~Aslam]%
       {Badruddin\thanks{email: \texttt{badr.physamu@gmail.com}} and O.~P.~M. ~Aslam\\
      Department of Physics, Aligarh Muslim University, Aligarh-202002, India.}

\pubyear{2012}
\volume{00}
\pagerange{\pageref{firstpage}--\pageref{lastpage}}


\date{Received --- ; accepted ---}

\maketitle

\label{firstpage}

\begin{abstract}
We utilize solar rotation average geomagnetic index {\it ap} and various solar wind plasma and field parameters for four solar cycles 20-23. We perform analysis to search for a best possible coupling function at 27-day time resolution. Regression analysis using these data at different phases of solar activity (increasing including maximum/decreasing including minimum) led us to suggest that the time variation of interplanetary electric potential is a better coupling function for solar wind-magnetosphere coupling. We suspect that a faster rate of change in interplanetary electric potential at the magnetopause might enhance the reconnection rate and energy transfer from the solar wind into the magnetosphere. The possible mechanism that involves the interplanetary potential fluctuations in influencing the solar wind-magnetosphere coupling is being investigated. 
\end{abstract}
\section{Introduction}\label{s:intro}
In the area of solar-terrestrial physics, one of the key problems is to investigate the mechanism of energy transfer from the solar wind into the magnetosphere. It is generally believed that the basic parameter leading to geomagnetic disturbances is the southward component of the interplanetary magnetic field (${-\it B}_z$) and/or the duskward component of the interplanetary electric field ${\it E}_y = -{\it V}\times {\it B}_z$ (see e.g. \citealt{{1961Phys. Rev. Lett. 6, 47D}};  \citealt{{1975J. Geophys. Res. 80, 4204B}};  \citealt{{1998Planet. Space Sci. 46, 1015B}}; \citealt{{2008J. Atmos. Sol. Terr. Phys. 70, 245G}}; Badruddin \& Singh 2009; Kane 2010; Alves, Echer \& Gonzalez 2011; Singh \& Badruddin 2012; Yermolaev et al. 2012 and references therein). In spite of the success of the so called Dungey mechanism some effort (e.g. Murayama 1982; Gupta \& Badruddin 2009; Joshi et al. 2011) has gone into looking for other parameters that might correlate better with geomagnetic activity. Geomagnetic activity being influenced by irregularities in the solar wind and interplanetary magnetic field, and enhanced dynamic pressure (Murayama 1982; Srivastava \& Venkatakrishnan 2002; Xie et al. 2008; Dwivedi, Tiwari \& Agrawal 2009; Singh \& Badruddin 2012) have been suggested. But a unique relationship is still lacking which may ultimately lead to understand the intensity of geomagnetic disturbances under different solar wind conditions. 
\section{Analysis and results}\label{s:fonts}
We have divided a complete solar cycle into two parts; (i) increasing including maximum and (ii) decreasing including minimum phases. We consider the time variations in different parameters at solar rotation time scale, both during increasing including maximum and decreasing including minimum phases of solar cycle 23. These time variations in various interplanetary plasma/field parameters ({\it V, B}, {\it B}$_{z}$, {\it E}$_{y}$, {\it BV}/1000 and {\it BV}$^{2}$) are compared with time variations of geomagnetic parameter {\it ap} during increasing including maximum phase (see Figure 1). From these figures, it appears that, at this time scale, {\it BV}$^{2}$ follows in a better way the time variation of {\it ap}, as compared to other parameters {\it V, B}, {\it B}$_{z}$, {\it E}$_{y}$ and {\it BV}/1000 considered here.
\begin{figure}
\centerline{\includegraphics[height=4.2cm,width=12cm]{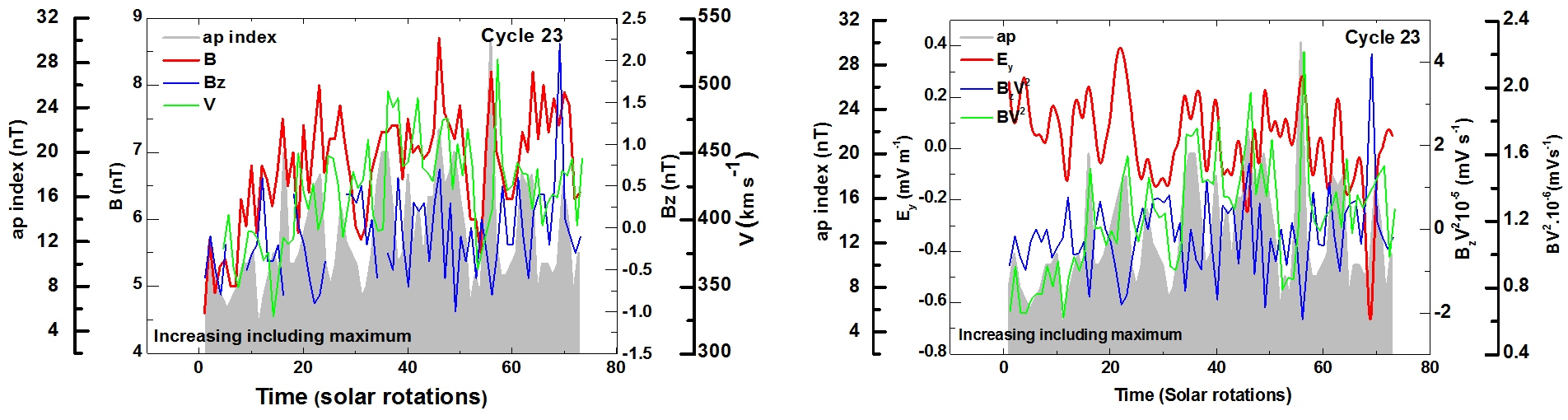}}
\caption{Time variation (27-day solar rotation average) of various interplanetary plasma/field parameters with geomagnetic {\it ap} index during increasing including maximum phase of solar cycle 23.}
\end{figure}

Figure 1 provides only qualitative idea about the time variations of various solar wind parameters as compared to geomagnetic parameter {\it ap} during increasing including maximum phase of solar cycle 23. Therefore, a quantitative analysis has been done by linear regression method, not only during increasing including maximum phase of solar cycle 23, but also during similar phases of solar cycles 20, 21, and 22. The rate of change of the {\it ap} index with various plasma/field parameters ($\Delta {\it I}/\Delta {\it P}$) obtained from the linear regression method and the corresponding correlation coefficients between them are tabulated in Table 1, during increasing including maximum phases of solar cycles 20, 21, 22 and 23. From Table 1 we observe that, out of various plasma/field parameters, the best correlation is found with the rate of change of interplanetary electric potential ({\it BV}$^{2}$) with time, consistently during increasing including maximum phases of all four solar cycles 20, 21, 22 and 23.  
\begin{continuedfigure}
\renewcommand\thefigure{2}
\centerline{\includegraphics[height=3.75cm,width=12cm]{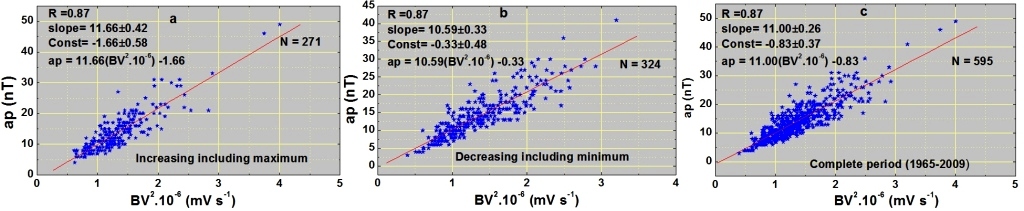}}
\caption{Scatter plot and best-fit linear curve between {\it ap} and {\it BV}$^2$ during (a) increasing including maximum phases, (b) decreasing including minimum phases and (c) complete cycles 20-23 combined.}
\end{continuedfigure}

\begin{table}
\caption{Rate of change of {\it ap} with various solar wind parameters ($\Delta {\it I}/\Delta {\it P}$) and correlation coefficient ({\it R}) during increasing including maximum phases of solar cycles 20, 21, 22 and 23.}
\begin{center}
\begin{tabular}{c c p{0.4cm} c p{0.4cm} c p{0.4cm} c p{0.4cm}} 
\hline
Parameters &\multicolumn{2}{c}{Solar cycle 20} & \multicolumn{2}{c}{Solar cycle 21}& \multicolumn{2}{c}{Solar cycle 22} &\multicolumn{2}{c}{Solar cycle 23}\\

&$\Delta {\it I}/\Delta {\it P}$ & {\it R} & $\Delta {\it I}/\Delta {\it P}$& {\it R} & $\Delta {\it I}/\Delta {\it P}$ & {\it R} & $\Delta {\it I}/\Delta {\it P}$ & {\it R} \\
\hline

{\it B} (nT) &3.44 &0.57&2.81 &0.55&3.99&0.74&3.77& 0.68\\

{\it B}$_z$ (nT)&$-3.44 $&$-0.42$&$-1.32$&$-0.13$&$-3.08$&$-0.28$&$-2.65$ &$-0.29$\\ 


{\it V} (km s$^{-1}$)&0.08&0.69&0.10&0.78&0.12&0.77&0.10&0.77\\ 

{\it P} (n Pa)&8.08&0.35&5.41&0.54&7.43&0.80&7.62&0.53\\

{\it E}$_{y}$ &9.38&0.61&2.35&0.10&5.76&0.24&6.27&0.29\\	
(mV m$^{-1}$)\\

{\it BV}/1000 &7.3&0.76&6.35&0.76&7.38&0.87&7.0&0.81\\ 
(mV m$^{-1}$)\\

{\it BV}$^{2}.10^{-6}$ &$1.14E{-5}$&0.77&$1.14E{-5}$&0.82&$1.85E{-5}$&0.90&$1.2E{-5}$&0.84\\  
(mV s$^{-1}$)\\

{\it VB}$^{2}$ &$6.9E{-4}$&0.71&$4.8E{-4}$&0.70&$4.98E{-4}$&0.84&$6.3E{-4}$&0.79\\  
 
\hline

\end{tabular}
\label{table_wide}
\end{center}
\end{table}

Similar correlative analysis during decreasing including minimum phases of solar cycles 20, 21, 22 and 23 was also done and the values of the rate of change of geomagnetic index {\it ap} with various parameters ($\Delta {\it I}/\Delta {\it P}$) and the corresponding correlation coefficients were calculated. We found, in this case also, that out of all parameters considered here, the rate of change of the interplanetary electric field ({\it BV}$^{2}$) with time best correlates with the {\it ap} during decreasing including minimum phases of almost all the four cycles 20, 21, 22 and 23. However, fluctuations in the interplanetary electric field occur in a much shorter time scale. Therefore we extended our analysis to 3-hourly resolution data especially for increasing including maximum and decreasing including minimum phases of solar cycle 23. We find that, at this shorter time scale too, the correlation coefficient ({\it R}) between {\it ap} and {\it BV}$^{2}$ is high ({\it R} = 0.69 during increasing including maximum phase and {\it R} = 0.72 for decreasing including minimum phase).  

The scatter plots of geomagnetic parameter {\it ap} with {\it BV}$^{2}$ for increasing including maximum phase [Figure 2 (a)], decreasing including minimum phase [Figure 2 (b)] and combined periods [Figure 2 (c)] with values of slope ($\Delta {\it I}/\Delta {\it P}$) and {\it R} are shown in Figure 2. From these figures we can give a best-fit relationship between {\it ap} and {\it BV}$^{2}$  as; $ap = 1.10\times10^{-5} (BV^{2})- 0.83$. This relation may be useful for estimating the geomagnetic activity level from the values of solar/interplanetary parameters {\it B} and {\it V}.

\section{Conclusions}

The time variability of interplanetary electric potential at the magnetopause appears to be an important parameter for solar wind-magnetosphere coupling. When this variability in interplanetary potential is sufficiently large, it appears to increase the reconnection rate between the solar wind and terrestrial magnetosphere, significantly increasing the geoefficiency of the solar wind. We suspect that, in addition to the duskward electric field, enhanced fluctuations in the interplanetary electric potential at the magnetopause is the most likely additional effect that leads to enhanced coupling between the solar wind and the terrestrial magnetosphere, significantly increasing the geoeffectiveness of the solar wind. However, this hypothesis needs to be tested with high time resolution measurements of in situ interplanetary data.

\label{lastpage}
\end{document}